\documentstyle[epsfig]{mn}

\newif\ifAMStwofonts

\ifoldfss
  \newcommand{\rmn}[1] {{\rm #1}}

  \ifCUPmtlplainloaded \else
    \NewTextAlphabet{textbfit} {cmbxti10} {}
    \NewTextAlphabet{textbfss} {cmssbx10} {}
    \NewMathAlphabet{mathbfit} {cmbxti10} {} 
    \NewMathAlphabet{mathbfss} {cmssbx10} {} 
  \fi
  \ifAMStwofonts
    \ifCUPmtlplainloaded \else
      \NewSymbolFont{upmath} {eurm10}
      \NewSymbolFont{AMSa} {msam10}
      \NewMathSymbol{\upi}     {0}{upmath}{19}
      \NewMathSymbol{\umu}     {0}{upmath}{16}
      \NewMathSymbol{\upartial}{0}{upmath}{40}
      \NewMathSymbol{\leqslant}{3}{AMSa}{36}
      \NewMathSymbol{\geqslant}{3}{AMSa}{3E}

      \let\leq=\leqslant 
       
    \fi
  \fi
\fi 

\ifnfssone
  \newmathalphabet{\mathit}
  \addtoversion{normal}{\mathit}{cmr}{m}{it}
  \addtoversion{bold}{\mathit}{cmr}{bx}{it}
  \newcommand{\rmn}[1] {\mathrm{#1}}

  \def\textbfit{\protect\txtbfit}
  
  \long\def\txtbfit#1{{\fontfamily{cmr}\fontseries{bx}\fontshape{it}%
    \selectfont #1}}

  \newmathalphabet{\mathbfit} 
  \addtoversion{normal}{\mathbfit}{cmr}{bx}{it}
  \addtoversion{bold}{\mathbfit}{cmr}{bx}{it}
  \newmathalphabet{\mathbfss} 
  \addtoversion{normal}{\mathbfss}{cmss}{bx}{n}
  \addtoversion{bold}{\mathbfss}{cmss}{bx}{n}
  \ifAMStwofonts
    \ifCUPmtlplainloaded \else
      \UseAMStwoboldmath
      \makeatletter
      \new@mathgroup\upmath@group
      \define@mathgroup\mv@normal\upmath@group{eur}{m}{n}
      \define@mathgroup\mv@bold\upmath@group{eur}{b}{n}
      \edef\UPM{\hexnumber\upmath@group}
      \new@mathgroup\amsa@group
      \define@mathgroup\mv@normal\amsa@group{msa}{m}{n}
      \define@mathgroup\mv@bold\amsa@group{msa}{m}{n}
      \edef\AMSa{\hexnumber\amsa@group}
      \makeatother
      \mathchardef\upi="0\UPM19
      \mathchardef\umu="0\UPM16
      \mathchardef\upartial="0\UPM40
      \mathchardef\leqslant="3\AMSa36
      \mathchardef\geqslant="3\AMSa3E

      \let\leq=\leqslant 

    \fi
  \fi
\fi 

\ifnfsstwo
  \newcommand{\rmn}[1] {\mathrm{#1}}

  \def\textbfit{\protect\txtbfit}
  
  \long\def\txtbfit#1{{\fontfamily{cmr}\fontseries{bx}\fontshape{it}%
    \selectfont #1}}

  \DeclareMathAlphabet{\mathbfit}{OT1}{cmr}{bx}{it}
  \SetMathAlphabet\mathbfit{bold}{OT1}{cmr}{bx}{it}
  \DeclareMathAlphabet{\mathbfss}{OT1}{cmss}{bx}{n}
  \SetMathAlphabet\mathbfss{bold}{OT1}{cmss}{bx}{n}
  \ifAMStwofonts
    \ifCUPmtlplainloaded \else
      \DeclareSymbolFont{UPM}{U}{eur}{m}{n}
      \SetSymbolFont{UPM}{bold}{U}{eur}{b}{n}
      \DeclareSymbolFont{AMSa}{U}{msa}{m}{n}
      \DeclareMathSymbol{\upi}{0}{UPM}{"19}
      \DeclareMathSymbol{\umu}{0}{UPM}{"16}
      \DeclareMathSymbol{\upartial}{0}{UPM}{"40}
      \DeclareMathSymbol{\leqslant}{3}{AMSa}{"36}
      \DeclareMathSymbol{\geqslant}{3}{AMSa}{"3E}

      \let\leq=\leqslant 

    \fi
  \fi
\fi 

\ifCUPmtlplainloaded \else
  \ifAMStwofonts \else 
    \def\upi{\pi}
    \def\umu{\mu}
    \def\upartial{\partial}
  \fi
\fi

\newcommand{\etal}{et~al.}
\newcommand{\scinum}[2]{#1\times 10^{#2}}
\newcommand{\iras}{{\it IRAS\/}}
\newcommand{\irasjy}{\iras{} $1.2$-Jy}
\newcommand{\bootes}{Bo\"otes}
\newcommand{\hmpc}[1]{#1 h^{-1} \mbox{Mpc}}
\newcommand{\rvol}{\mbox{$r_0$}}
\newcommand{\ab}{A_B}
\newcommand{\apj}{ApJ}
\newcommand{\apjs}{ApJS}
\newcommand{\mnras}{MNRAS}
\newcommand{\aj}{AJ}
\newcommand{\nat}{Nature}
\newcommand{\physrep}{Phys.~Rep.}
\newcommand{\wb}{{\sc wall builder}}
\newcommand{\vf}{{\sc void finder}}
\newcommand{\sect}[1]{\S \ref{#1}}
\newcommand{\phn}   {\phantom{0}}  
\newcommand{\phmgt} {\phantom{$>$}}  
\newcommand{\fig}[1]{Fig.~\ref{#1}}
\newcommand{\tab}[1]{Table~\ref{#1}}
\newcommand{\equ}[1]{Equ.~\ref{#1}}

\title[ORS voids vs.\ {\it IRAS\/} voids]
      {A case devoid of bias: ORS voids vs.\ \textbfit{IRAS\/} voids}
\author[El-Ad \& Piran]
       {H.~El-Ad$^{1,2}$
         \thanks{e-mail: \protect\verb+helad@cfa.harvard.edu+} and
         T.~Piran$^{2,3,4}$
         \thanks{e-mail: \protect\verb+tsvi@nikki.fiz.huji.ac.il+}\\
        $^1$Harvard-Smithsonian Center for Astrophysics, 60 Garden
        Street, Cambridge, MA 02138\\
        $^2$Racah Institute of Physics, The Hebrew University,
        Jerusalem 91904, Israel\\
        $^3$Astronomy Department, Columbia University, 550 West 120th
        Street, New York, NY 10027\\
        $^4$Department of Physics, New York University, New York, NY
        10003}

\pagerange{\pageref{firstpage}--\pageref{lastpage}}

\pubyear{1999}

\begin{document}

\maketitle

\label{firstpage}

\begin{abstract}
  
  We present a comparison between the voids in two nearly all sky
  redshift surveys: the ORS and the \irasjy. While the galaxies in
  these surveys are selected differently and their populations are
  known to be biased relative to each other, the two void
  distributions are similar. We compare the spatial distribution of
  the two void populations and demonstrate the correlation between
  them. The voids also agree with regard to the overall void
  statistics -- a filling factor of $\approx 0.45$ of the volume, an
  average void diameter $ \bar{d}_{\rmn{void}} \approx \hmpc{45} $ and
  an average galaxy underdensity in the voids $ \bar{\delta} \approx
  -0.9 $. Our measurements of the underdensities of the voids in the
  two surveys enable us to estimate the relative bias in the voids
  between optical and \iras{} samples. We find $(b_{\rmn{opt}} / b_{\it
    IRAS})^{\rmn{void}} \approx 1$, showing that on average there is
  little -- or no -- biasing between the two void populations.

\end{abstract}

\begin{keywords}
       cosmology: observations --
       galaxies: clustering --
       large-scale structure of the universe.
\end{keywords}

\section{Introduction}

The Optical Redshift Survey (ORS; Santiago \etal{} 1995) 
and the Infrared Astronomical Satellite (\iras) $1.2$-Jy redshift
survey \cite{fi95} comprise an interesting data set pair: both are
nearly all-sky surveys (excluding the Galactic plane region: $|b| <
20\degr$ for the ORS, $|b| < 5 \degr$ for the \iras) and as such are
the densest, wide angle, three-dimensional samples currently available
for study (at least until the forthcoming release of the \iras{}
PSC$z$ Redshift Survey, complete to 0.6-Jy -- see Saunders \etal{} 
1998).  While sampling approximately the same volume, the two surveys
differ vis-\`{a}-vis the galaxy populations they probe: \iras{} and
optical catalogues are each compiled using different selection
criteria.  As a result, \iras{} galaxies are biased relative to
optically selected galaxies -- namely, they are less clustered and
underrepresented in cores of galaxy clusters \cite{st92}.

As such, the two samples were already compared in several studies,
e.g., the study of galaxy clustering 
and morphological segregation in the ORS \cite{sh96} and the derivation
of the ORS-predicted velocity field \cite{jeb98}, compared with the \irasjy{}
gravity field \cite{fi95}. These works focused on properties derived
from the distribution of galaxies in the surveys.  In this study, we
attempt a different approach, now comparing the distributions of voids
in the two surveys.

Voids are the most prominent feature of the large-scale structure of
the universe, indeed occupying more than a half of the volume. Thus
they are natural candidates for any quantitative large-scale structure
study. We have already derived a void catalogue for the \irasjy{}
survey \cite{epd97}, and in this paper we rederive a similar catalogue
and present a new void catalogue for the ORS, using a suitably
modified version of the \vf{} algorithm \cite{ep97}.

But in addition to this additional void catalogue, the most
interesting aspect of this paper is perhaps the comparison between the
two void populations: if the galaxies in one survey are biased
relative to the other, how does this affect the distribution of the
voids? Being almost empty, and using a code which does its best at
trying not to depend on the details of the galaxy distribution, voids
could prove to be a relatively bias-free statistical probe.

\begin{table*}
\begin{minipage}{\textwidth}
\caption{ORSd sub-catalogue parameters}
\label{ORSdtable}
\begin{tabular}{@{}lccccccccc@{}}
Sample & Volume & $N_{\rmn{gal}}$ &$\alpha$&$\beta$& $r_*$     & $N_{\rmn{gal}}$  & $n_1(5)$ & $n_1(50)$ & \\
       &Fraction& (total)&      &       &(km s$^{-1}$)&(used)& \multicolumn{2}{c}{($10^{-2} h^{-3} \mbox{Mpc}^3$)} & \raisebox{2mm}[-2mm]{$\frac{\bar{n}_{{\it IRAS}_j}}{\bar{n}_{\it IRAS}}$} \\
\multicolumn{1}{c}{(1)}&(2)&(3)& (4) & (5)   & (6)  & (7) & (8) & (9) & (10) \\
\hline
ESGC & 0.14 & 1203 & 0.50 & 5.40 &    10545 & 684 & $    20.4$ & 0.68 & 1.233\\
ESO  & 0.33 & 1639 & 0.39 & 3.22 &\phn 6425 & 696 & $\phn 6.8$ & 0.25 & 0.760\\
UGC  & 0.53 & 1903 & 0.33 & 2.75 &\phn 4085 & 648 & $\phn 9.6$ & 0.18 & 1.025\\
\hline
\end{tabular}
\end{minipage}
\end{table*}

In this paper we derive in a consistent manner and compare void
catalogues of the ORS and the \irasjy{} surveys. The paper is
structured as follows. In \sect{sampsect} we describe the redshift
catalogues we use, and in \sect{algosect} we briefly review the \vf{}
code and detail the modifications incorporated in order to analyze a
survey with a non-isotropic selection function as the ORS. We then
introduce the void catalogues (\sect{catsect}) and note some of the
familiar voids we identify (\sect{cosmosect}). In \sect{discsect} we
compare the two void catalogues. Finally, in \sect{sumsect} we
summarize our main conclusions.

\section{The samples}
\label{sampsect}

\begin{figure}
  \psfig{bbllx=105, bblly=220, bburx=335, bbury=707, clip=,
    file=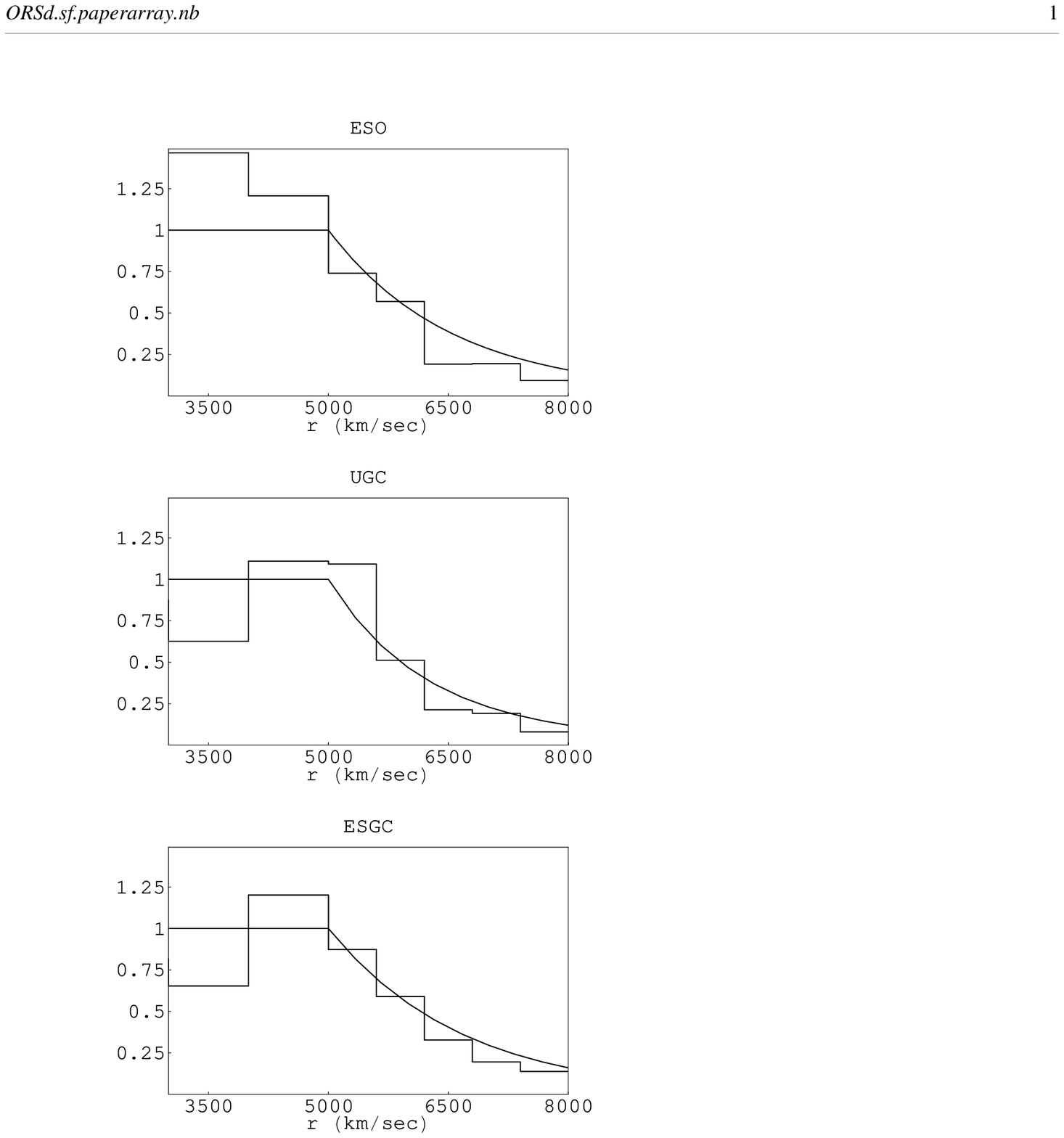, width=\linewidth}
  \caption[]
  {ORSd sub-catalogues radial density profiles for our $\hmpc{50}$
    semi--volume-limited samples extending to $\hmpc{80}$.  The smooth
    curves are the calculated selection functions, normalized
    according to each catalogue's number density $n_1(50)$ (see
    \tab{ORSdtable}, column~9). }
  \label{sfsfig}
\end{figure}

From the original redshift catalogues, we construct two
semi--volume-limited samples with the same geometry: a sphere
extending out to $\hmpc{80}$ with the volume-limited region comprising
the inner $\rvol = \hmpc{50}$. The Galactic plane is cut out of our
samples eliminating the $|b| < 20 \degr$ region, as we are limited by
the wider ZOA of the ORS; hence our samples extend over 66 per cent of the
skies. The volume examined is $ \scinum{1.41}{6} h^{-3} \mbox{Mpc}^3 $.

The ORS catalogue \cite{ors95} contains over 8000 galaxies with
redshifts, drawn from three sources -- the UGC, ESO and ESGC
catalogues. We choose to work with the diameter-limited ORSd
sub-sample, as its sky coverage is wider than that of the
magnitude-limited ORSm sub-sample (ORSm does not include the ESGC
strip). After applying our geometrical cuts and volume-limiting, we
end up with 2028 galaxies (\tab{ORSdtable} provides a break down of
the galaxy counts per sub-catalogue: column~3 details the original
ORSd catalogues, and column~7 details our final sample). The catalogue
contains seven $z$-collapsed clusters. Note that since extinction
corrections are properly taken into account, the volume limited region
is not a perfect sphere -- at directions where extinction is not
negligible, the volume limited region is shallower having a depth
$\rvol / 10^{0.2 \gamma \ab}$.  Here $\ab(l,b)$ is the extinction in
the given direction and passband.  $\gamma$ is the extinction
correction parameter, for the ORSd being $\gamma_d$, the fractional
decrease in isophotal diameter with extinction. We used $\gamma_d =
0.6$ throughout \cite{ors96}.

The various ORSd selection functions were derived as outlined in
Santiago \etal{} \shortcite{ors96}. We use a parameterized form for
the selection functions \cite{ya91}:
\begin{equation}
  \phi(r) = \left(\frac{r}{r_s}\right)^{-2\alpha}
            \left(\frac{r_*^2 + r^2}{r_*^2 + r_s^2}\right)^{-\beta}
\end{equation}
where $\phi(r < r_s) = 1$ and $\alpha$, $\beta$ and $r_*$ are free
parameters whose best-fit values for our specific samples \cite{bs98}
are given in \tab{ORSdtable}.

The \iras{} catalogue contains 5321 galaxies complete to a flux limit of
1.2-Jy \cite{fi95}. The sample we used, selected as explained above,
has 1362 galaxies. It is important to note that the two catalogues are
not independent -- about half of the \iras{} galaxies also appear in
ORSd, and we have carefully taken this into account in our statistical
analysis (\sect{discsect}). All of the analysis is performed in $z$ space.

\begin{table*}
\begin{minipage}{\textwidth}
\caption{Locations \& properties of ORSd voids}
\label{voidtable}
\begin{tabular}{@{}cccccrrrcccl@{}}
ORSd &            & Equivalent & Total & \multicolumn{4}{c}{Location of Centre} & Void & & Void & \\
Void & Confidence & Diameter & Volume & \multicolumn{4}{c}{(Supergalactic Coordinates)} & Under- & \irasjy{} & Fit & Identification \\
\cline{5-8}
Index & Level & {[$\hmpc{}$]} & {[$h^{-3}\mbox{KMpc}^3$]} & \multicolumn{1}{c}{$r$} & \multicolumn{1}{c}{$X$} & \multicolumn{1}{c}{$Y$} & \multicolumn{1}{c}{$Z$} & density & Counterpart & $\eta$ & \\
(1) & (2) & (3) & (4) & (5) & \multicolumn{1}{c}{(6)} & \multicolumn{1}{c}{(7)} & \multicolumn{1}{c}{(8)} & (9) & (10) & (11) & \multicolumn{1}{c}{(12)} \\
\hline
1& $>$0.99 & 64.9&   142.9 &51.7 &$-10.5$&$-35.0$&$-36.5$&$-0.87$&5+9&0.38& EPdC6+7 \\        
2& $>$0.99 & 62.9&   130.4 &56.5 &$-29.3$&$-48.2$&$  3.2$&$-0.87$& 1 &0.58& EPdC5 (Sculptor)\\
3& $>$0.99 & 37.7&\phn28.2 &41.9 &$ -8.7$&$-29.0$&$ 29.0$&$-0.87$& 8 &0.49& SV2 \\            
4& $>$0.99 & 53.8&\phn81.7 &54.2 &$-26.0$&$ 34.7$&$ 32.5$&$-0.93$& 3 &0.45& \\                
5& $>$0.99 & 59.7&   111.6 &57.7 &$ 37.1$&$ 41.6$&$ 14.8$&$-0.85$& 2 &0.49& tip of T1 \\      
6&\phmgt0.98&37.2&\phn26.9 &30.5 &$ -2.9$&$ 28.8$&$-9.9$&$-0.94$&4+11&0.31& Local--Coma \\    
7&\phmgt0.98&35.6&\phn23.6 &55.5 &$-40.9$&$ 31.7$&$-20.1$&$-0.90$&10 &0.48& \\                
8&\phmgt0.98&44.2&\phn45.2 &55.3 &$  9.8$&$ 29.2$&$ 45.9$&$-0.86$& 6 &0.42& CfA2n \\          
9&\phmgt0.98&46.5&\phn52.6 &63.6 &$ 15.5$&$-47.1$&$ 39.8$&$-0.89$& 7 &0.60& CfA2s, tip of T2\\
\hline
\end{tabular}
\end{minipage}
\end{table*}

\section{Modifications in the {\sevensize\bf VOID FINDER} algorithm}
\label{algosect}

The \vf{} code used here to derive the void catalogues has been
described in detail elsewhere \cite{ep97}. Briefly, the code covers
voids using overlapping spheres, iteratively working its way starting
from voids containing the largest spheres. Subsequent iterations
identify new voids containing smaller spheres and improve the coverage
of previously identified voids. Since voids need not be completely
empty, an initial phase (\wb) is used in order to filter out isolated
galaxies which are allowed to be in the voids. Corrections are applied
in the code in order to handle the observational selection function
$\phi(r)$. In a magnitude-limited sample, as we probe deeper we
observe a smaller fraction of the galaxy distribution, hence the
significance of finding an empty sphere declines with distance. The
code corrects for this observational effect by weighing galaxies
(during the initial phase) and spheres (during the construction of the
voids) according to their distance.

The ORS is more complicated to analyze than the previous surveys we
have worked with (SSRS2s and \irasjy{}), since the usage of three
different catalogues in its making (and the required extinction
corrections) result in a non-isotropic selection function.  We
modified the \vf{} code in order to take this into account by
appropriately weighing the galaxies and the spheres used to compose
the voids. Each galaxy is assigned a weight \cite{ors96}:
\begin{equation}
  W_i = \frac{1}{\bar{n}_j \phi_j(\bmath{r}_i)}
        \frac{\bar{n}_{{\it IRAS}_j}}{\bar{n}_{\it IRAS}}
  \label{weiequ}
\end{equation}
where $\bar{n}_j$ is the mean number density of ORS galaxies in
sub-catalogue $j$ in which galaxy $i$ happens to be located;
$\bar{n}_{{\it IRAS}_j}$ is the mean number density of \iras{}
galaxies inside that sub-catalogue; and $\bar{n}_{\it IRAS}$ is the
total mean number density of \iras{} galaxies (see
\tab{ORSdtable}, column~10).

The selection function of each sub-catalogue $\phi_j(r_i)$ is usually
just a function of the distance $r$, but in order to take extinction
into account we adjust it:
\begin{equation}
  \phi_j(\bmath{r}_i) = 10^{0.2 \gamma \ab} \phi_j(r_i) 
\end{equation}
where $\ab(l,b)$ are the direction-dependent absorption coefficients
\cite{bh82}.

Consequently, we calculate a weight for each sphere considered to be a
part of a void by volume averaging over the weights of points within
each sphere:
\begin{equation}
  \bar{W}({\bmath{r}_{\rm centre}},d) =
    \left(\frac{4 \pi}{3} d^3\right)^{-1}
    \int_{\rm volume} W({\bmath{r}})
\end{equation}
for a sphere centred on ${\bmath{r}_{\rm centre}}$ with radius $d$.

In order to estimate the underdensity of the ORSd voids we derive
$n_{1j}(r_0)$, the selection function based galaxy number density for
each sub-catalogue $j$ up to $r_0$. The prescribed number density is
$n_{1j}(r_s)$ where $r_s = \hmpc{5}$ (see \tab{ORSdtable}, column~8),
and we calculate $n_{1j}(r_0) = n_{1j}(r_s) \phi_j(r_0)$ (column~9).
The actual galaxy number density for $r < r_0$ in our samples agrees
well with the $n_{1j}(r_0)$ values, except for ESOd where the actual
number density is significantly higher due to the presence of four
nearby clusters (Doradus, Hydra, Centaurus and Fornax). See
\fig{sfsfig}.

For the purpose of calculating the void underdensities: if a void
extends over several sub-catalogues, we derive the underdensity in
each part of the void separately, and then volume-average the partial
underdensities; and if a void extends beyond $r_0$ we weigh the
galaxies in it using the relevant catalogue's selection function. Note
that since the calculation is done separately in each sub-catalogue
$j$ relative to $n_{1j}$, the weight now is simply $1 / \phi_j$ and no
relative density corrections (as in \equ{weiequ}) are required. So the
underdensity of a void with volume $V$ containing galaxies at
locations $\bmath{r}_i$ is:
\begin{equation}
  \frac{\delta \rho}{\bar{\rho}} =
    \sum_j f_j
    \left( \frac{\sum_i \phi_j^{-1}(\bmath{r}_i)}{V} / n_{1j}(50) - 1 \right)
  \label{udequ}
\end{equation}
where $f_j$ is the fraction of the void that happens to be in
sub-catalogue $j$.

\begin{figure*}
  \psfig{bbllx=108, bblly=349, bburx=500, bbury=627, clip=,
    file=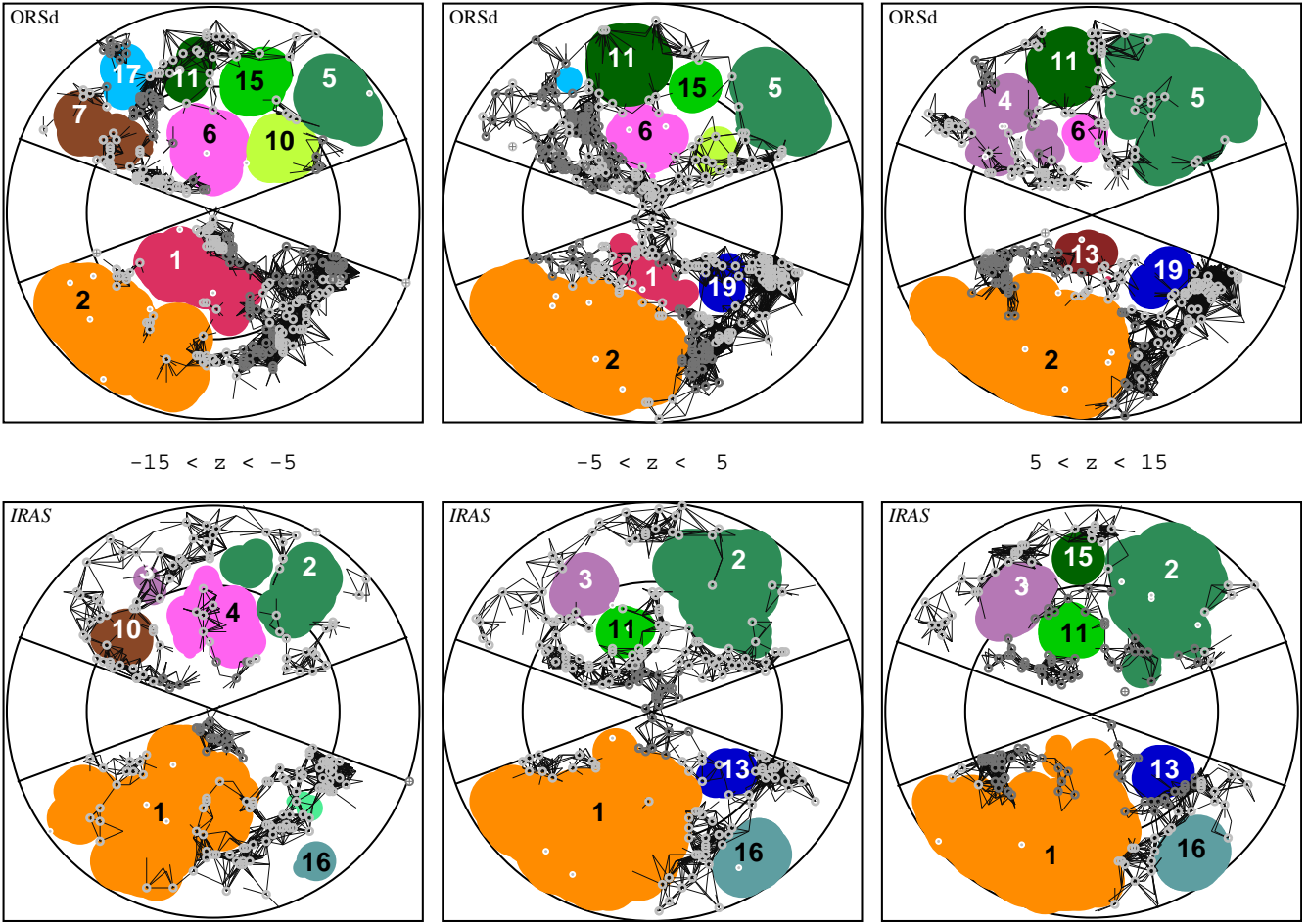, width=\linewidth}
  \caption[]
  {ORSd vs. \iras{} slices: for each survey, we present 3 consecutive
    slices, above and below the supergalactic plane covering the range
    $- 15 < Z_{\rmn{SG}} < 15$. The geometry of our samples is
    indicated by the outer circles, marking the boundary of the
    samples at $\hmpc{80}$, and the inner circles, marking the
    volume-limited region at $\rvol = \hmpc{50}$. The excluded ZOA at
    $|b| < 20 \degr$ cuts across the samples along $ Y_{\rmn{SG}} = 0
    $.  The coloured areas mark the intersection of the
    three-dimensional voids with the centres of the indicated slices.
    Voids are numbered as in \tab{voidtable} (but the table details
    only the significant voids); the smaller the void's index, the
    more significant it is.  Galaxies are marked with dots, and
    neighboring wall galaxies (i.e., galaxies not allowed to be in
    voids) are connected with lines. }
  \label{slcfig}
\end{figure*}

To estimate the voids' statistical significance we use our usual {\it
  confidence level} \cite{ep97}:
\begin{equation}
  p(d) = 1 - \frac{N_{\rm Poisson}(d)}{N(d)} 
\end{equation}
where $ N_{\rm Poisson}(d) $ is the number of voids in a Poisson
distribution that contain a sphere whose diameter is $d$, and $N(d)$
is the same quantity for an actual survey. Poisson distributions are
constructed using the same luminosity functions and extinction
coefficients as the survey they correspond to.  Our quoted confidence
level should not be confused with the usual $\sigma$ grade; as such,
it is a rather conservative grade since it does not take into account
the total volume of a void, but rather only the size of the largest
sphere that fits into it. Our $p$ is based on this aspect of the voids
since it is the size of the largest sphere within a void that triggers
a void's initial identification by the \vf.

\section{The void catalogues}
\label{catsect}

\subsection{ORS\lowercase{d}}
\label{orssub}

The \wb{} identified 1909 (94 per cent) of the galaxies as wall
galaxies which may not reside in voids. Of the remaining 119 galaxies,
100 were found to be in voids (see \tab{comptable}).

We identified 19 voids in the ORSd for which $p > 0$; of these 9 have
$p > 0.95$, and we list these in \tab{voidtable}: Column (1)
identifies the voids with index numbers.  Column (2) indicates $p$,
the confidence level. Column (3) lists the diameters of equal-volume
spheres; the volumes are tabulated in column (4). Column (5) lists the
distance to the void centres, and the centres locations are detailed,
in supergalactic coordinates, in column (6)--(8). Column (9) lists the
void underdensities. Column (10) indicates the matching void(s) in our
\iras{} void catalogue (see \sect{irassub}), and column (11) measures
the fit between the corresponding voids (see \equ{equeta} in
\sect{discsect}).  Finally, column (12) identifies some of the
familiar voids we identify (see \sect{cosmosect}).

\begin{table*}
\begin{minipage}{\textwidth}
\caption{ORSd vs.\ \iras{} galaxy and void statistics}
\label{comptable}
\begin{tabular}{@{}lcccccccc@{}}
Sample &\multicolumn{4}{c}{Galaxies}& Number & Void   &$\bar{\delta}$& $\bar{d}_{\rmn{void}}$ \\
\cline{2-5}
       &Total &Wall &Non-wall &Void &of Voids&Fraction&          &[$\hmpc{}$]\\
(1)    & (2)  & (3) &  (4)    & (5) & (6)    & (7)    & (8)      & (9) \\
\hline
ORSd & 2028& 1909 (94.1\%)& 119 (5.9\%)&    100 (4.9\%)& 10+9 & 0.46+0.08&$-0.89$&49\\
\iras& 1362& 1260 (92.5\%)& 102 (7.5\%)&\phn 71 (5.2\%)& 11+5 & 0.43+0.05&$-0.88$&44\\
\hline
\end{tabular}
\end{minipage}
\end{table*}

The 9 ORSd voids with $p > 0.95$ occupy 46 per cent of the survey's
volume; an additional 8 per cent are occupied by the remaining 10
voids, bringing the void filling factor to $0.54$.  The average
equivalent diameter of the 9 significant voids is
$\bar{d}_{\rmn{ORSd}} = \hmpc{49}$ and the average underdensity in
these voids is $\bar{\delta}_{\rmn{ORSd}} = -0.89$.

\subsection{\textbfit{IRAS\/}}
\label{irassub}

In the \iras{}, $92.5$ per cent of the galaxies were located in walls,
and 5 per cent in the voids (see \tab{comptable}). The \vf{}
identified 16 voids with $p > 0$ in the \iras; of these, the initial
11 correspond to the above mentioned 9 significant ORSd voids (see
\tab{voidtable}, column~10). These 11 voids occupy 43 per cent of the
volume (with the additional 5 voids occupying 5 per cent), with
$\bar{d}_{\it IRAS} = \hmpc{44}$ and $\bar{\delta}_{\it IRAS} =
-0.88$.

\section{Cosmography}
\label{cosmosect}

Two of the significant voids (4 and 7; and perhaps also 5) and most of
the $p < 0.95$ voids identified here are new and are not listed in the
literature.  Familiar voids which were already listed elsewhere are
indicated in column (12) of \tab{voidtable}.

Voids 1--3 are all located within the volume of space probed by the
Southern Sky Redshift Survey (SSRS; da Costa \etal{} 1998). 
We have already compiled a void catalogue of the southern Galactic cap
$m_B \leq 15.5$ edition of this survey (SSRS2s), and we identify
void~1 with EPdC voids~6+7; void~2 corresponds to EPdC void~5
\cite{ep97}. Void~2 is also known as the Sculptor void, and was
identified in an earlier SSRS paper \cite{dc88} as SV3. Void~3 was
identified in that paper (see Table~1 there) as SV2.  Void~9 is
pointed out in the recently published south Galactic cap $m_B \leq
15.5$ CfA survey (CfA2s; Huchra, Vogeley \& Geller 1999). 
Void~8 corresponds to the large void found in the $26\fdg5 < \delta <
32\fdg5$ slice of the (north) CfA2 survey \cite{lgh86}; this void is
also identified (marked V4) in Table~1 of Saunders \etal{}
\shortcite{sau91}.

The voids listed by Tully \shortcite{tu86} are mostly beyond the range
of our sample, but it is likely that voids 5 and 9 are the nearby tips
of Tully voids 1 and 2, respectively (see Table~1 in the above
referenced paper).  Tully's {\em Local Void} \cite{tu87} is defined to
cover the region closer than $\hmpc{30}$ in the approximate direction
(in Galactic coordinates) $|b| < 60\degr$ and $0\degr < l < 90\degr$.
We find in this direction voids 4, 8 (north of the Galactic plane), 3
and 9 (south of the Galactic plane) -- though they all lie deeper than
$\hmpc{30}$.

The cosmographical tour of Strauss \& Willick \shortcite{st95}
mentions several other voids identified here: the void indicated to
lie between the Local and Coma superclusters is void~6. The void
beyond the Virgo cluster is void~11 and the void in the foreground of
the Perseus-Pisces supercluster is void~19. The latter two voids have
$p < 0.95$ and thus are not listed in \tab{voidtable}, but they can be
viewed in \fig{slcfig}.

\begin{figure}
  \psfig{bbllx=240, bblly=331, bburx=370, bbury=461, clip=,
    file=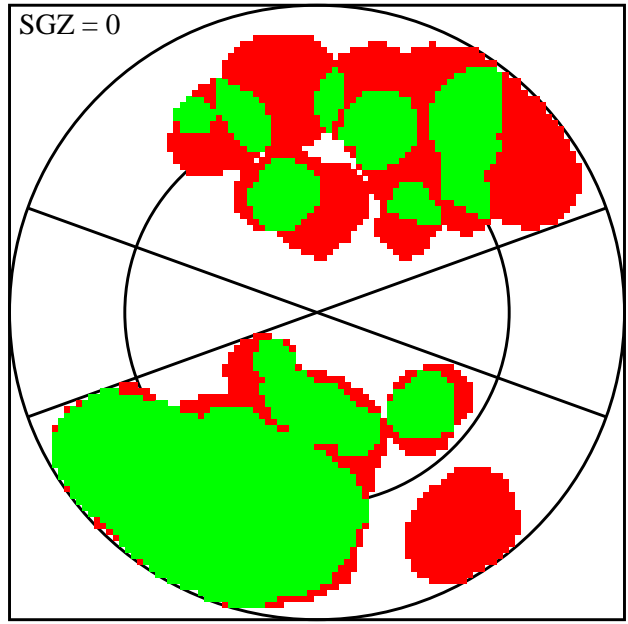, width=\linewidth}
  \caption[]
  { Overall overlap measure of ORSd and \iras{} voids in the
    supergalactic plane $Z_{\rmn{SG}} = 0$. Green marks overlap, red
    marks no overlap. The overlap score in this slice is $\eta =
    0.43$.}
  \label{etafig}
\end{figure}

\section{Discussion}
\label{discsect}

The two void images are similar, although the two galaxy samples are
quite different. In \fig{slcfig} we present two sets of slices
covering the supergalactic plane and slices immediately above and
below it, all together spanning \mbox{$- 15 < Z_{\rmn{SG}} < 15$}.

Beyond the visual impression, we quantify the spatial similarity
between the two void populations by deriving $\eta$, the ratio between
the overlapping volume of the two void distributions and their union:
\begin{equation} 
  \eta = \frac{V_1 \cap V_2}{V_1 \cup V_2} 
  \label{equeta} 
\end{equation} 
where $V_1$ represents the volumes occupied by voids in one
distribution, and $V_2$ is the same quantity for the other
distribution. In case of exact overlap we have $\eta = 1$; for two
random distributions the expected overlapping volume is simply $f_1
f_2$, the product of the two void filling factors. E.g., if $f_1 = f_2
= 0.5$ the expected score for a pair of random samples is $\eta =
1/3$. On the other hand, if 80 per cent of the volumes overlap we
would get $\eta = 2/3$. See \fig{etafig} for an illustration of the
void overlap in the supergalactic plane.

Serendipitously, the details for our case happen to be quite similar 
to the above example: $f_{\rmn{ORSd}} = 0.54$ and $f_{\it IRAS} = 
0.48$. The overlap score is $\eta_{{\rmn{ORSd-}{\it IRAS}}} = 0.58$. 
In contrast, for random ORSd- and \iras{}-like samples where half of
the random \iras{} locations are used in the random ORSd distributions
we get $\eta_{\rmn{rnd0.5}} = 0.19 \pm 0.04$; for completely unmatched
sample pairs the score is $\eta_{\rmn{rnd}} = 0.13 \pm 0.02$.  The
theoretical expectation for $\eta_{\rmn{rnd}}$, based on the filling
factors of the random samples, is $0.10 \pm 0.02$, and the small
actual excess over it is likely due to the geometrical constraints of
the sample.

The above test measures the overall correlation between two void
distributions without trying to match individual voids.  As we
identified (by eye) the corresponding voids in the two surveys
(\tab{voidtable}, column~10), we can also measure how well do the
individual voids overlap. We do this by deriving $\eta$ values for a
void from one of the samples and its counterpart(s) in the other
sample. We report $\eta$ values for the ORSd--\iras{} pairs in
\tab{voidtable}, column (11). The average value is $\bar{\eta} =
0.47$. We can also give this score a more intuitive interpretation, by
converting it to $d_{\rmn{misfit}}(\eta)$, the distance (in fractions
of a diameter units) at which one would need to place two identical
spheres in order to get a specific value of $\eta$. Hence if there is
exact overlap $\eta = 1$ and $d = 0$. If there is no overlap $\eta =
0$ and $d > 1$. A good reference point is at $d = 0.5$ (two
identical spheres misplaced by one radius), for which $\eta = 5/27$. 
The average result we got for the ORSd--\iras{} pairs corresponds to
$d_{\rmn{misfit}}(\bar{\eta}) = 0.25$.

Note that difference in void volumes is a contributing factor in the
derivation of the (mis)fit scores. E.g., for two spheres at the same
center but with one radius being $0.57$ of the other we would get the
same $\eta$ as for identical spheres one radius apart. As on average
the ORSd voids are somewhat bigger than the \iras{} voids, we can
quantify the contribution of this factor to the misfit score: it
translates to a base misfit value of $d_{\rmn{misfit(volume)}} \approx
0.1$.

In addition to the spatial correlation, the two distributions are also
similar with regard to the average void properties -- the total void
filling factor, the average void diameter $\bar{d}_{\rmn{void}}$ and
the average void underdensity $\bar{\delta}$ (see \tab{comptable}). We
find the later similarity to be of special interest: estimates of the
relative bias between optical and \iras{} samples based on the
distribution of galaxies find $(b_{\rmn{opt}} / b_{\it
  IRAS})^{\rmn{gal}} \approx 1.5$ \cite{lnp90,jeb98}. However, as the
galaxy underdensity in the voids in both surveys is $\bar{\delta}
\approx -0.9$, the \vf{} analysis shows that on average there is
practically no biasing in the voids: $(b_{\rmn{opt}} / b_{\it
  IRAS})^{\rmn{void}} = 1$.

The only other work we are aware of which compared optical and \iras{}
galaxies in voids examined the \bootes{} void \cite{dsh90}. There it
was found that the density contrast of \iras{} galaxies within the
\bootes{} sphere is roughly equal to the (optical) upper limit for
that region \cite{koss87}. In this work we examine a distribution of
voids, and use many more galaxies ($\approx 100$, compared to 12
\iras{} galaxies in the \bootes{}); still, our result of little -- or
no -- biasing between optical and \iras{} galaxies in the voids is
consistent with the \bootes{} result.

\section{Summary}
\label{sumsect}

In this paper we present a comparison between two void distributions.
These distributions sample the same volume of space, but were derived
using the \vf{} code from two different galaxy samples -- chosen
optically (ORSd) and by the \iras{}. The 9 significant voids we find
in the ORSd match very well the locations of their \iras{}
counterparts, and our overlap/union ($\eta$) test shows a correlation
significantly in excess of random. Combined with our previous analysis
of the SSRS2s sample, we now have 3 different void catalogues all
showing similar void properties, including the filling factor, average
equivalent diameter and underdensity.

In all our samples so far voids are limited by the boundaries of the
surveys -- in this paper, by the ZOA and the limited depth
($\hmpc{80}$); and in the SSRS2s by the narrow declination span
(37\fdg5). In order to overcome this limitation we intend to further
extend our void catalogues using deeper (LCRS -- Shectman \etal{} 
1996) and wider (CfA2s -- Huchra, Vogeley \& Geller 1999) samples. 

The fact that we find practically no biasing in the voids
$(b_{\rmn{opt}} / b_{\it IRAS})^{\rmn{void}} = 1$ indicates that voids
may be a relatively bias-free environment. As such, they comprise an
attractive target with which one can examine different cosmological
models, and we intend to explore this possibility using $N$-body
simulations.

\noindent {\bf Acknowledgments.}
We are indebted to Basilio Santiago for his help in unraveling the
mysteries of the ORS. We thank Ofer Lahav, Myron Lecar, David Meiri
and Sune Hermit for helpful discussions and comments. HE was supported
by a Smithsonian Predoctoral Fellowship.

\bsp 

\label{lastpage}


\begin{thebibliography}{}
  
\bibitem[\protect\citename{Baker \etal\ }
  1998]{jeb98} Baker J. E., Davis M., Strauss M. A., Lahav O.,
  Santiago B. X., 1998, \apj, 508, 6

\bibitem[\protect\citename{Burstein \& Heiles }%
  1982]{bh82} Burstein D., Heiles C., 1982, \aj, 87, 1165

\bibitem[\protect\citename{da Costa \etal\ }%
  1988]{dc88} da Costa L. N., \etal, 1988, \apj, 327, 544

\bibitem[\protect\citename{da Costa \etal\ }%
  1998]{dc98} da Costa L. N., \etal, 1998, \aj, 116, 1

\bibitem[\protect\citename{de Lapparent, Geller \& Huchra }%
  1986]{lgh86} de Lapparent V., Geller M. J., Huchra J. P., 1986,
  \apj, 302, L1

\bibitem[\protect\citename{Dey, Strauss \& Huchra }%
  1990]{dsh90} Dey A., Strauss M. A., Huchra J., 1990, \aj, 99, 463

\bibitem[\protect\citename{El-Ad, Piran \& da Costa }
  1997]{epd97} El-Ad H., Piran T., da Costa L. N., 1997, \mnras, 287,
  790

\bibitem[\protect\citename{El-Ad \& Piran }
  1997]{ep97} El-Ad H., Piran T., 1997, \apj, 491, 421

\bibitem[\protect\citename{Fisher \etal\ }
  1995]{fi95} Fisher K. B., Huchra J. P., Strauss M. A., Davis M.,
  Yahil A., Schlegel D., 1995, \apjs, 100, 69

\bibitem[\protect\citename{Hermit \etal\ }
  1996]{sh96} Hermit S., Santiago B. X., Lahav O., Strauss M. A.,
  Davis M., Dressler A., Huchra J. P., 1996, \mnras, 283, 709

\bibitem[\protect\citename{Huchra, Vogeley \& Geller }%
  1999]{hvg99} Huchra J. P., Vogeley M. S., Geller M. J., 1999, \apjs,
  121, 287

\bibitem[\protect\citename{Kirshner \etal\ }%
  1987]{koss87} Kirshner R. P., Oemler A. Jr., Schechter P. L.,
  Shectman S. A., 1987, \apj, 314, 493

\bibitem[\protect\citename{Lahav, Nemiroff \& Piran }%
  1990]{lnp90} Lahav O., Nemiroff R. J., Piran T., 1990, \apj, 350,
  119

\bibitem[\protect\citename{Santiago \etal\ }
  1995]{ors95} Santiago B. X., Strauss M. A., Lahav O., Davis M.,
  Dressler A., Huchra J. P., 1995, \apj, 446, 457

\bibitem[\protect\citename{Santiago \etal\ }
  1996]{ors96} Santiago B. X., Strauss M. A., Lahav O., Davis M.,
  Dressler A., Huchra J. P., 1996, \apj, 461, 38

\bibitem[\protect\citename{Santiago }%
  1998]{bs98} Santiago B. X., 1998, private communication

\bibitem[\protect\citename{Strauss \etal\ }%
  1992]{st92} Strauss M. A., Davis M., Yahil A., Huchra J. P., 1992,
  \apj, 385, 421

\bibitem[\protect\citename{Saunders \etal\ }%
  1991]{sau91} Saunders W., \etal, 1991, \nat, 349, 32

\bibitem[\protect\citename{Saunders \etal\ }%
  1998]{sau98} Saunders W., \etal, 1998, in Proc. of the XXXII Moriond
  Astrophysics Meeting, {\it in press}

\bibitem[\protect\citename{Shectman \etal\ }
  1996]{lcrs} Shectman S. A., Landy S. D., Oemler A., Tucker D. L.,
  Lin H., Kirshner R. P., Schechter P. L., 1996, \apj, 470, 172

\bibitem[\protect\citename{Strauss \& Willick }
  1995]{st95} Strauss M. A., Willick J. A., 1995, \physrep, 261, 271

\bibitem[\protect\citename{Tully }%
  1986]{tu86} Tully R. B., 1986, \apj, 303, 25

\bibitem[\protect\citename{Tully }%
  1987]{tu87} Tully R. B., 1987, Nearby Galaxies Atlas.  Cambridge
  University Press, Cambridge

\bibitem[\protect\citename{Yahil \etal\ }%
  1991]{ya91} Yahil A., Strauss M. A., Davis M., Huchra J. P., 1991,
  \apj, 372, 380

\end{thebibliography}
\end{document}